\documentclass[journal]{IEEEtran}

\usepackage{graphicx}
\usepackage{latexsym}
\usepackage{amsfonts}
\usepackage{amssymb}
\usepackage{psfrag}
\usepackage{subfigure}
\usepackage{comment,cancel,float}
\usepackage{amsmath,epsfig,epstopdf,color,lineno,makeidx,booktabs,multirow}
\usepackage[normalem]{ulem}
\usepackage{cases}
 
 \usepackage[numbers, square,comma,sort&compress]{natbib}

\title{Robust Data Detection for the Photon-Counting Free-Space Optical System with Implicit CSI Acquisition and Background Radiation Compensation}

\author{Tianyu Song, \IEEEmembership{Student Member,~IEEE} and~Pooi-Yuen~Kam,~\IEEEmembership{Fellow,~IEEE} 
\thanks{This work was supported by the Singapore MoE AcRF Tier 2 Grant MOE2010-T2-1-101. }
\thanks{The authors are with the Department of Electrical \& Computer Engineering, National University of Singapore (e-mail: \{song.tianyu, elekampy\}@nus.edu.sg).}
}

\begin{document}

\maketitle

\begin{abstract}
Since atmospheric turbulence and pointing errors cause signal intensity fluctuations and the background radiation surrounding the free-space optical (FSO) receiver contributes an undesired noisy component, the receiver requires accurate channel state information (CSI) and background information to adjust the detection threshold.  
In most previous studies, for CSI acquisition, pilot symbols were employed, which leads to a reduction of spectral and energy efficiency; and an impractical assumption that the background radiation component is perfectly known was made.
In this paper, we develop an efficient and robust sequence receiver, which acquires the CSI  and the background information implicitly and requires no knowledge about the channel model information. 
It is robust since it can automatically estimate the CSI and background component and detect the data sequence accordingly.
Its decision metric has a simple form and involves no integrals, and thus can be easily evaluated. 
A Viterbi-type trellis-search algorithm is adopted to improve the search efficiency, and a selective-store strategy is adopted to overcome a potential error floor problem as well as to increase the memory efficiency.
To further simplify the receiver, a decision-feedback symbol-by-symbol receiver is proposed as an approximation of the sequence receiver.
By simulations and theoretical analysis, we show that the performance of both the sequence receiver and the symbol-by-symbol receiver, approach that of detection with perfect knowledge of the CSI and background radiation, as the length of the window for forming the decision metric increases. 
\end{abstract}


\section{Introduction}
~\IEEEPARstart{F}{ree}-space optical (FSO) communications provide high data rate transmission with higher security and higher flexibility compared with conventional wireless communications. 
However, FSO signals are attenuated when transmitted through air by atmospheric phenomena such as rain, fog and snow that reduce the range of the system and deteriorate the quality of the transmission. 
Also, atmospheric turbulence and pointing errors cause fluctuations in the intensity of the received signal and further degrade the link performance. 
Additionally, the background radiation constitutes an undesired component to the received signal which also fluctuates due to the environmental uncertainty.    

Due to the complexity of phase and frequency modulation, intensity modulation (IM) is used for most current FSO  systems. 
There are mainly two  types of detection methods: direct detection (DD) and photon counting (PC), both of which require accurate knowledge of the channel state information (CSI) and background radiation for reliable data detection.
In \cite{Song2014Arobust, Song2014GlobeCom}, we have surveyed the literature and considered the detection issues for IM/DD systems. 
In particular, we have applied the generalized likelihood ratio test (GLRT) principle to design a robust receiver that performs implicit CSI acquisition and automatic threshold adjustment. 
The receiver is shown to be able to achieve the error performance of a receiver with perfect CSI.
In this paper, we will focus on IM/PC systems. 
In \cite{Wilson2005JSAC, Wilson2005TOC}, pulse position modulation (PPM) is used, which does not require CSI and the background information for data detection but has less spectral efficiency compared to the IM.  
With the  assumption that the receiver perfectly knows the background information and the channel model information (CMI), i.e., the statistical distribution of the channel gain, in \cite{Schober2008PCTWC}, a maximum likeliohood (ML) sequence detection (MLSD) method for the photon-counting system  is proposed.
To simplify the implementation, a fast search algorithm called multi-symbol detection (MSD) algorithm, is also proposed in \cite{Schober2008PCTWC}.
With this MSD algorithm, the receiver performs block-by-block detection and the implementation complexity per symbol detection is reduced to O($\log(L)$) operations, where $L$ denotes the  block length. 
However, the requirement of the accurate CMI and background information can hardly be satisfied in practice.
Later,  a generalised MLSD (GMLSD) receiver, which requires no CMI and has a simpler form, is proposed in \cite{Uysal2010GMLSD} where the same MSD algorithm as that in \cite{Schober2008PCTWC} for implementation is suggested.
It works with neither CSI nor CMI, but requires the knowledge of background radiation, which still limits its applicability in time-varying environments. 
In addition, for the receiver in \cite{Uysal2010GMLSD}, an undesired error floor problem is observed.

The GLRT principle is a very powerful approach for receiver design in the presence of unknown channel parameters. 
Having applied it to the IM/DD system in \cite{Song2014Arobust, Song2014GlobeCom}, we apply it here to the IM/PC system and obtain a robust sequence receiver which can automatically and continuously estimate the CSI and the background radiation and detect the data sequence accordingly.  
A Viterbi-type trellis-search algorithm similar to that we have introduced in \cite{Song2014Arobust, Song2014GlobeCom}, which reduces the search complexity to a level that is independent of the observation window length, is adopted for efficient implementation.  
Besides, we use a selective-store strategy to overcome the error floor problem observed in \cite{Uysal2010GMLSD}. 
To further reduce the implementation complexity, a decision-feedback (DFB) symbol-by-symbol receiver is developed. 
To distinguish our two receivers from others, we call our sequence receiver the GLRT sequence receiver and our DFB receiver the GLRT DFB receiver. 
The simulation and analytical results show that as the observation window length increases, both our sequence receiver and DFB receiver can achieve the \emph{Genie Bound}, which is defined as the bit error probability (BEP) of the receiver with perfect CSI and background information.
Additionally, we slightly simplify the GMLSD receiver in \cite{Uysal2010GMLSD} and simulate it using our Viterbi-type trellis-search algorithm and selective-store strategy. 
The result shows that the error floor can be completely avoided.
We also simplify the GMLSD receiver in \cite{Uysal2010GMLSD} to a DFB receiver. 
We call the original receiver in  \cite{Uysal2010GMLSD}  the GMLSD sequence receiver, and call our simplification the GMLSD DFB receiver. 
Particular comparisons and discussions will be given in corresponding sections and the advantages of our GLRT (sequence and DFB) receivers  over the existing receivers will be explained.

The remaining parts of this paper are organized as follows. 
A mathematical system model of the FSO photon counting system perturbed by atmospheric turbulence, pointing errors and background radiation is briefly reviewed in section II. 
In section III, we briefly illustrate three existing receivers, i.e. the ideal receiver with perfect CSI and background information, the MLSD receiver introduced in \cite{Schober2008PCTWC}, and the GMLSD sequence receiver introduced in \cite{Uysal2010GMLSD}.  
We also give a brief discussion of the three receivers. 
In section IV,   we present our GLRT sequence receiver and the Viterbi-type trellis-search algorithm as well as our selective-store strategy.
 The GLRT DFB receiver is introduced in section V, and the GMLSD DFB receiver is developed as a special case of the GLRT DFB receiver.
The performance results of all the receivers and discussions are given in  section VI.
In the last section, our conclusions are drawn. 
The common italic font letters in this paper, such as $r$, $m$ and $L$, are used to denote scalar quantities;
the bold non-italic font lower-case letters, such as $\mathbf{r}$ and $\mathbf{m}$, are used to denote vectors and the  bold non-italic font upper-case letters, such as $\mathbf{A}_{r}$ and $\mathbf{A}_{dm}$, are used to denote memory arrays.

\section{System Model}
\subsection{Signal Model}
At each time $k$, the received signal $r(k)$ is a discrete Poisson random variable with probability mass function (PMF)  \cite{Schober2008PCTWC, Uysal2010GMLSD, Wilson2005JSAC, Wilson2005TOC}
\begin{align}
&P(r(k)|m(k),h,n_s,n_b)=  \nonumber \\
&\qquad  \qquad  \frac{(n_s m(k) h + n_b)^{r(k)}  \exp(-(n_s m(k) h + n_b))}{r(k)!},
\label{eq:sig_model}
\end{align}
where $h$ denotes the channel gain, $n_s$ and $n_b$ are the mean count parameters due to the transmitted signal and the background radiation, respectively. 
Notation $m(k)$ is used to denote the transmitted data bit, which takes on either the value ``0'' or ``1'' with equal probability.
Here, we define $n_r = hn_s$ as the the effective count parameters due to the received signal and in later parts of this paper, $n_r$ and $hn_s$ are used interchangeably. 
Consequently, $P(r(k)|m(k),h,n_s,n_b)$ and $P(r(k)|m(k),n_r,n_b)$ are used interchangeably.

\subsection{Channel Model}

As in \cite{Farid2007Outage}, the channel gain $h$ is formulated as  $h=h_p h_a h_l$, where $h_p$, $h_a$ and $h_l$ are used to denote geometric spread and pointing errors, atmospheric turbulence, and path loss, respectively.
In \cite{Schober2008PCTWC}, log-normal distribution is adopted to model $h_a$ for weak turbulence, and Gamma-Gamma distribution for moderate to strong turbulence and the negative exponential distribution for strong turbulence. 
Since in \cite{Andrews2001Mathe}, it has been shown that the Gamma-Gamma distribution can nicely fit the channel fading statistics of all turbulence regimes, in this paper, we only consider $h_a$ is a Gamma-Gamma distributed random variable, and the \emph{probability density function} (pdf) of $h_a$ is
\begin{multline}
p_{h_a}(h)=\frac{2(\alpha\beta)^{(\alpha+\beta)/2}}{\Gamma(\alpha)\Gamma(\beta)}
h^{(\alpha+\beta)/2-1} \\
\times K_{\alpha-\beta} \left(2\sqrt{\alpha\beta h}\right),\ h>0 ,
\label{eq:pdf_gg}
\end{multline}
where $K_{a}(\cdot)$ is the modified Bessel function of the second kind, and $1/\beta$ and $1/\alpha$ are the variances of the small and large scale eddies, respectively. 
Pointing error influence on an FSO system is discussed in  \cite{Farid2007Outage} and \cite{Deva2009PointingE}, and we here use the model in \cite{Farid2007Outage} where the pdf of $h_p$ is given as 
\begin{align}
p_{h_p}(h)=\gamma^2 h^{\gamma^2-1}/A_0^{\gamma^2},\ 0<h<A_0. 
\label{eq: pdf_pointing}
\end{align}
Parameters $A_0$ and $\gamma$ are constants, and further details can be found  in \cite{Farid2007Outage}.
In \cite{Farid2007Outage, Song2014Arobust}, the path loss $h_l$ is considered as a deterministic variable.
This is rigorously wrong since $h_l$ depends highly on the distance between the transmitter and the receiver and the atmospheric condition which is definitely time-varying.
However, if we only consider a short range of time, e.g. on the order of seconds, $h_l$ can be regarded as a constant.
Since we only consider the receiver side signal-to-noise ratio (SNR) which is defined later in \eqref{eq:snr}, we can incorporate $h_l$ into $h_a$ which amounts to setting $h_l=1$. 
Then for a turbulent channel with pointing errors, the channel gain is $h=h_{a} h_p$, and its pdf can be derived by performing 
\begin{align}
p_h(h)=\int_{0}^\infty \frac{1}{|a|}  p_{h_a}(a)p_{h_p}\left(\frac{h}{a}\right) da,\ h>0,
\label{eq:p_h}
\end{align}
where $p_{h_a}(a)$ is the pdf of $h_a$, and can be found in \cite{Farid2007Outage}.
It should be emphasized that \eqref{eq:p_h} is not an always-correct expression for the pdf of $h$ since we artificially set $h_l$ as 1. 
If a receiver highly depends on the knowledge of $p_h(h)$, using \eqref{eq:p_h} is inappropriate.
In our work, we will show that our receiver does not depend on  $p_h(h)$, and \eqref{eq:p_h} is only used to present some numerical results.

As pointing errors are considered as well as atmospheric turbulence in our paper, the scintillation index (SI), which is defined as the normalised variance of the irradiance fluctuations due to atmospheric turbulence  \cite{Mudge2011SI, Andrews2001Mathe}, is 
\begin{align}
\text{SI}=  \frac{\mathbb{E}[(h_a )^2] }{ \mathbb{E}^2[h_a ]}-1,
\end{align}
but not $\mathbb{E}[h^2]/\mathbb{E}^2[h]-1$.
Since we use the Gamma-Gamma fading model, the SI can also be calculated according to $\text{SI}= \alpha^{-1}+\beta^{-1}+(\alpha \beta)^{-1}$\cite{Andrews2001Mathe}.
The received signal-to-noise ratio (SNR), which is defined as the ratio of the squared expected mean of the information bearing component to the total variance of the received signal  \cite{gagliardi1995optical,Uysal2010GMLSD}, is  
\begin{align}
\mathbf{SNR} =  \frac{(n_s\mathbb{E}[h])^2}{(2n_s\mathbb{E}[h]+4n_b)}.
\label{eq:snr}
\end{align}
Specifically, if pilot symbols are required, for example, there are $P$ pilot symbols and $D$ data symbols in every data packet, the effective SNR is given by
\begin{align}
\mathbf{SNR}^e =\frac{P+D}{D}\mathbf{SNR}.
\end{align}

Since the time scales of the fading processes are of the order of $10^{-3}$s to $10^{-2}$s \cite{Xie2011EffICC}, which is far larger than the bit interval ($\approx 10^{-10}$s for multi-Gbps systems), $h$ is considered to be constant over a large number ($>10^{5}$) of transmitted data symbols. 
In general, the FSO system is exposed to ambient light, which is random and time-varying, hence the assumption made in \cite{Wilson2005TOC, Wilson2005JSAC, Schober2008PCTWC, Uysal2010GMLSD} that $n_b$ equals 39 constantly and is known at the receiver side, is not realistic in practice. 
According to the experimental results reported in \cite{khatoon2011channel, moreira1997optical}, the photocurrent due to background radiation is commonly in the kilo-hertz region.
For some special source, e.g., the Fluorescent lamps geared by electronic ballasts, the photocurrent fluctuates in the mega-hertz region.    
Thus, it is more reasonable to assume that the background radiation strength (both natural or artificial) is time-varying but keeps coherent during intervals of the order of $10^{-7}$s - $10^{-6}$s. 
Thus, with this assumption, the effective count parameter due to background radiation has a coherence time roughly on the order of  $10^{-7}$s -$10^{-6}$s.
Thus, the channel coherence length $L_c$, which is defined as the number of data symbols over which $h$ and $n_b$ can be regarded as constants, is of the order of $10^3$.

It should be emphasized that for our proposed GLRT receivers, the observation window length $L_w$ is much shorter than $L_c$, i.e., $L_w \ll L_c$.
Thus, within the observation window with length $L_w$, we can treat $n_s$, $n_b$ and $h$ (including $n_r$) as unknown but non-random constant parameters.
In our receiver design introduced in sections IV and V, the values of $n_s$ and $n_r$, and the distribution of $h$ are not required. 
We only use the distribution of $h$ and choose appropriate parameter values to run simulations and to show numerical results in section VI.

\section{Existing Receivers}
\subsection{The Ideal Receiver}
The ideal receiver is considered as a benchmark for the performance analysis of all other receivers in this paper.  
It is assumed to accurately have all the information, i.e., the accurate values of $n_b$ and $n_r$,  for detection. 
When $n_r \neq 0$, i.e., $P(r(k)|1, n_r, n_b ) \neq 0$, the decision rule is given by 
\begin{align}
\frac{P(r(k)|0, n_r, n_b ) }{P(r(k)|1, n_r, n_b ) } 
\mathop{\gtrless} \limits^{\hat m(k) = 0}_{\hat m(k) = 1} 1,
\label{eq:PCSI_1}
\end{align}
where $\hat m(k)$ denotes the decision on $m(k)$.  
When $n_b=0$, obviously, we have
\begin{align}
P(r(k)|0, n_r, n_b ) = 
\left\{
\begin{array}{rll}
& 1  &, r(k)=0\\
& 0 &,  \text{elsewhere}\\
\end{array}
\right.  .
\end{align}
Thus, the decision rule when $n_b=0$ is 
\begin{align}
\hat m(k) = 
\left\{
\begin{array}{rll}
& 0  &, r(k)=0\\
& 1 &,  \text{elsewhere}\\
\end{array}
\right.  .
\end{align}
When $n_b>0$, by substituting the signal PMF into \eqref{eq:PCSI_1} and simplifying it, the decision rule is reduced to
\begin{align}
r(k)  \mathop{\gtrless} \limits^{\hat m(k) = 1}_{\hat m(k) = 0} n_r / \ln \left(1+ \frac{n_r }{n_b} \right).
\label{eq:receiver_ideal}
\end{align}

The average bit error probability (BEP) over all possible channel states is given  by 
\begin{align}
&P(e|n_s, n_b) \nonumber \\ 
=&  
 \int_0^\infty \frac{ P(e|0,h,n_s,n_b)+P(e|1,h,n_s,n_b)}{2} p_h(h) dh ,
\label{eq:bep_ideal}
\end{align}
where $P(e|0,h, n_s,n_b)$ is the probability that a ``0'' is transmitted but a ``1'' is the decision made by the receiver, conditioned on a given channel parameter combination $(h, n_s, n_b)$; similarly, $P(e|1,h,n_s, n_b)$ is the conditional probability that a ``1'' is transmitted but a ``0'' is detected.
The value of $P(e|0,h, n_s, n_b)$ and $P(e|1,h, n_s, n_b)$ can be evaluated by 
\begin{align}
&P(e|0,h, n_s, n_b)   
= P(r(k)  >  \tau ) = 1 - F( \tau ,n_b)
\label{eq:bep_ideal0}
\end{align}
and
\begin{align}
& P(e|1,h,n_s,n_b) 
=  P(r(i)  < \tau )  =  F( \tau,hn_s+n_b),
\label{eq:bep_ideal1}
\end{align}
where $F(k,\lambda)$ is the \emph{cumulative distribution function} (cdf) of a Poisson distribution with parameter $\lambda$.
In \eqref{eq:bep_ideal0} and \eqref{eq:bep_ideal1}, $\tau$ denotes the decision threshold, i.e., $\tau = \ln \left(1+ {n_r }/{n_b} \right)$.
The average BEP obtained from \eqref{eq:bep_ideal} is also referred to as the Genie Bound.

From \eqref{eq:bep_ideal} - \eqref{eq:bep_ideal1}, we see that the error probability is related to $n_s$, $n_b$ and the pdf of $h$. 
Unlike some other systems, the SNR value alone cannot determine the BEP here.
For example, if $n_r=50, n_b=25$, the corresponding SNR is 12.5 and BEP is $1.17\times10^{-4}$; if $n_r=100, n_b=150$,  the corresponding SNR is till 12.5 but BEP changes to $1.80\times10^{-4}$.

Also, we see that increasing the value of $n_b$ will decrease the SNR and thus degrade the system performance, even for the ideal receiver. 
For receivers using an inaccurate value of $n_b$, the error performance is further degraded. 
We will show that our proposed receivers can approach the ideal receiver's performance, but not completely cancel the background radiation.

\subsection{The MLSD Receiver}
A MLSD receiver for the MIMO photon-counting system has been introduced in \cite{Schober2008PCTWC}, and here we just introduce its SISO case. 
It assumes that at the receiver side the channel model information, together with the value of $n_b$ which is 39 constantly, is available perfectly. 
At each time $k$, we consider a subsequence $\mathbf{m}(k, L)$ of the immediate past $L$ transmitted data symbols given by $\mathbf{m}(k,L)=[m(k-L+1), ... , m(k)]$, where  $m(i)\in\{0, 1\}$, $\forall i$. 
Similarly,  $\mathbf{r}(k,L)=[r(k-L+1),  ... , r(k)]$ is used to denote the corresponding received signal subsequence. 
The MLSD receiver performs joint detection on the transmitted subsequence based on
\begin{align}
&\mathbf{\hat m}(k,L)
= \arg \max_{\mathbf{m}(k,L) }
P(\mathbf{r}(k,L)|\mathbf{m}(k,L),n_b) \nonumber \\
=&\arg \max_{\mathbf{m}(k,L) }\int_0^\infty \prod_{i=k-L+1}^k P(r(i)|m(i),h,n_b) p_h(h) dh. 
\label{eq:MLSD_1}
\end{align}
 By eliminating irrelevant terms in \eqref{eq:MLSD_1}, the decision rule is reduced to
\begin{align}
\mathbf{\hat m}(k,L) =  
 \arg \max_{\mathbf{m}(k,L) }  \lambda_0(\mathbf{m}(k,L)),
 \label{eq:receiver_mlsd}
 \end{align}
where  
 \begin{multline}
 \lambda_0(\mathbf{m}(k,L))  = 
\int_0^\infty (\frac{h n_s}{n_b} + 1)^{R_\mathrm{on}(\mathbf{  m}(k,L))}  \\
   \times \exp(-(n_s N_\mathrm{on}(\mathbf{  m}(k,L)) h  + n_b L)) p_h(h) dh
\label{eq:metric_mlsd}
\end{multline}
 denotes its decision metric for subsequence $ \mathbf{m}(k,L)$.
Notation $\mathbf{\hat m}(k,L)$ is used to denote the decision on $\mathbf{m}(k,L)$.
Quantities $N_\mathrm{on}(\mathbf{  m}(k,L))$ and $R_\mathrm{on}(\mathbf{  m}(k,L))$ are defined as
\begin{align}
N_\mathrm{on}(\mathbf{  m}(k,L))&=\sum_{i=k-L+1}^k m(i), 
\label{eq:Non_def}
\end{align}
and
\begin{align}
R_\mathrm{on}(\mathbf{  m}(k,L))&=\sum_{i=k-L+1}^k m(i)r(i).
\label{eq:Ron_def}
\end{align}
For simplicity of notation, we drop the dependence on $ \mathbf{  m}(k,L) $ for $N_\mathrm{on}(\mathbf{  m}(k,L))$ and $R_\mathrm{on}(\mathbf{  m}(k,L))$ and use $N_\mathrm{on}$ and $R_\mathrm{on}$ instead, but it should be emphasized that $N_\mathrm{on}$ and $R_\mathrm{on}$ are functions of $\mathbf{  m}(k,L)$. 
This MLSD receiver performs block-by-block detection using a multi-symbol detection (MSD) algorithm \cite{Schober2008PCTWC}.  
Blocks with length $L$ are considered.
To find the optimal MSD solution in practice, we first let 
 \begin{align}
 g(1) \geq g(2) \geq ... \geq g(L) 
 \label{eq:MLSD_sorting}
\end{align} 
denote the sorted values of $r(k)$, ordered from the largest to the smallest. 
Secondly, we define $G_\mathrm{on}(N_\mathrm{on})$ to be the sum of the $N_\mathrm{on}$  largest values of $r(k)$;
that is
 \begin{align}
 G_\mathrm{on}(N_\mathrm{on}) = \sum_{i=1}^{N_\mathrm{on}} g(i).
 \end{align}
To determine what value of $N_\mathrm{on}$ maximizes the MSD metric, \eqref{eq:metric_mlsd} must be evaluated for each $N_\mathrm{on} = 0, 1, . . .,L$, using its partner $R_\mathrm{on} = G_\mathrm{on}(N_\mathrm{on})$. 
According to \cite{Schober2008PCTWC}, the MSD sequence will correspond to the estimate $\hat N_\mathrm{on}$ which satisfies
\begin{align}
\hat N_\mathrm{on} = \arg \max \lambda_0( N_\mathrm{on}, G_\mathrm{on}).
\end{align}
By using the reverse mapping of the sorting associated with
\eqref{eq:MLSD_sorting}, the final decision $\mathbf{\hat m}(k,L)$ can be generated. 
Specifically, ones will be assigned to the indices corresponding to the   largest values of $r(k)$ in $\mathbf{r}$, and zeros will be assigned to the remaining $L - \hat N_\mathrm{on}$ elements of $\mathbf{\hat m}(k,L)$.

In summary, $\text{O}(L \log_2 L)$ operations are required for sorting, $\text{O}(L)$ to calculate $G_\mathrm{on}(N_\mathrm{on})$ for all $N_\mathrm{on}$, and $\text{O}(L)$ to perform the $L$ metric evaluations. 
Thus, the algorithm has an overall complexity of $\text{O}(L \log_2 L)$ operations per $L$ symbol decisions, or $\text{O}(\log_2 L)$ operations per symbol decision, and is only logarithmically dependent on $L$. 
This is a significant reduction relative to the complexity of a brute
force search, which is $\text{O}(2^L/L)$ on a per symbol decision basis.

\subsection{The GMLSD Receiver}
This GMLSD receiver is proposed in \cite{Uysal2010GMLSD}.
It assumes that the channel model information is unavailable at the receiver side; but the value of $n_b$ is 39 constantly and known by the receiver. 
This GMLSD receiver jointly decides on $\mathbf{m}(k, L)$ and $h$ that maximize $P(\mathbf{r}(k,L)|\mathbf{m}(k,L),h, n_b)$, which is the PMF of the received subsequence and is given by 
\begin{align}
P(\mathbf{r}(k,L)&|\mathbf{m}(k,L),h, n_s, n_b) \nonumber \\
&=\prod_{i=k-L+1}^k P(r(i)|m(i),h, n_s, n_b ) 
\end{align}
We use  $\hat h$ to denote the estimate  of $h$. 
For a given $\mathbf{m}$, the solution of equation
\begin{align}
\frac{d P (\mathbf{r}(k,L)|\mathbf{m}(k,L),h, n_s, n_b)}{dh} = 0,
\end{align}
which is given by 
\begin{align}
\hat{h}(\mathbf m(k,L))=\left( \frac{R_\mathrm{on}}{N_\mathrm{on}}-n_b \right) \frac{1}{n_s},
\end{align}
makes $P(\mathbf{r}(k,L)|\mathbf{m}(k,L),h,n_s, n_b)$ achieve its maximum  value.
By substituting $h=\hat h(\mathbf m(k,L))$ into $P(\mathbf{r}(k,L)|\mathbf{m}(k,L),h, n_s, n_b)$ and eliminating irrelevant terms, the decision metric becomes
\begin{align}
 \label{eq:metric_gmlsd_seq}
 \lambda_1(\mathbf{m}(k,L)) = \left( \frac{R_\mathrm{on} }{N_\mathrm{on} n_b} \right)^{R_\mathrm{on} }   \exp(-R_\mathrm{on} +n_bN_\mathrm{on} )
\end{align} 
 Similar to \eqref{eq:receiver_mlsd}, this GMLSD receiveris decision is  
 \begin{align}
 \mathbf{\hat m}(k,L) =  
 \arg \max_{\mathbf{m}(k,L) }  \lambda_1(\mathbf{m}(k,L)).
\label{eq:receiver_gmlsd_seq} 
 \end{align}
It also performs block-by-block detection  and uses the MSD algorithm introduced in \cite{Schober2008PCTWC} and briefly introduced in the previous subsection.

 \subsection{A Brief Summary and Discussion}
In implementation, the ideal receiver must have exact values of $h$, $n_s$ and $n_b$.  
However, in practice, without channel estimation, knowing these values is impossible. 
Using pilot symbols to estimate these parameters reduces bandwidth and energy efficiency, and thus obviously is not desirable. 
In this paper, we do not consider channel estimation with pilot symbols.
The performance of the ideal receiver is just referred to as a benchmark when analysing other receivers' performance.
The benefit of the MLSD receiver proposed in \cite{Schober2008PCTWC} is to obviate the need for accurate values of $h$.
Nevertheless, the evaluation of its decision metric which involves complicated integrals demands high computational capability of the receiver hardware. 
The search complexity of the MSD algorithm increases 
with the number of symbol decisions in the block.
As long blocks are preferred for better performance, a large complexity seems inevitable and receiver hardware with very high computational capability is a prerequisite. 
Besides, when performing block-by-block detection, using a large block length $L$ brings a long system delay.
Even if a powerful processor is available, the requirement that the exact distribution of $h$ and the exact value of $n_b$ be known makes the implementation impractical.   
The GMLSD receiver, introduced in \cite{Uysal2010GMLSD},  does not require the distribution of $h$ to be known.
Another benefit is that the evaluation of its decision metric can be easily performed since no integrals are involved.  
However, the undesired features due to the MSD algorithm, such as a high search complexity and a large system delay, still exist. 
Besides, an error floor problem is observed in \cite{Uysal2010GMLSD}.  
Furthermore, since the accurate value of $n_b$ is required, the performance of this GMLSD receiver may deteriorate when the receiver is exposed to time-varying ambient light. 

In later sections of this paper, we will develop new receivers, and propose new implementation methods to overcome all the problems mentioned above. 
 
\section{The GLRT Sequence Receiver}

\subsection{   Decision metric design based on the GLRT principle}

As discussed previously, the receiver is likely to be exposed to  time-varying, environmental ambient radiation, even though by employing a well-designed shade, an optical filter, or both, the ambient light component can be decreased. 
Nevertheless, a residual component with weak but time-varying intensity is inevitable.    
Since the value of $n_b$ is unavailable at the receiver side, the GLRT sequence receiver jointly decides on $\mathbf{m}(k, L)$, $n_r$ ($=hn_s$) and $n_b$ that maximize $P (\mathbf{r}(k,L)|\mathbf{m}(k,L),n_r, n_b)$.

At time $k$, the PMF of the received subsequence is  
\begin{align}
P(\mathbf{r}(k,L)|\mathbf{m}(k,L), n_r, n_b) = \prod_{i=k-L+1}^k P(r(i)|m(i),n_r, n_b ) 
\end{align}
For a given $\mathbf{m}$, the solution of the simultaneous equations 
\begin{align}
\frac{d P (\mathbf{r}(k,L)|\mathbf{m}(k,L),n_r, n_b)}{d n_r} = 0,  \label{eq:diff1}
\end{align}
\begin{align}
\frac{d P (\mathbf{r}(k,L)|\mathbf{m}(k,L),n_r, n_b)}{d n_b} = 0,  \label{eq:diff2}
\end{align}
makes $P (\mathbf{r}(k,L)|\mathbf{m}(k,L),n_r,n_b)$ achieve its maximum  value.
We first get the solution of \eqref{eq:diff1}, which is
\begin{equation}
n_r = \hat n_r(\mathbf m(k,L))=\left(\frac{R_\mathrm{on} }{N_\mathrm{on} }-n_b\right),
\label{eq:nr_estimate}
\end{equation} 
By substituting the solution $n_r = \hat n_r(\mathbf m(k,L))$ back into $P(\mathbf{r}(k,L)|\mathbf{m}(k,L),n_r , n_b)$, we have 
\begin{align}
&P(\mathbf{r}(k,L)|\mathbf{m}(k,L),\hat n_r, n_b)\nonumber \\   
=&  
\frac{ n_b ^{R_\mathrm{on} +R_\mathrm{off} }\exp(-n_bL)}  {\prod_{i=k-L+1}^k (r(i)!)} 
  \left( \frac{R_\mathrm{on} }{N_\mathrm{on} n_b}\right) ^{R_\mathrm{on} }   \exp(-R_\mathrm{on} +n_bN_\mathrm{on} ),
  \label{eq:GLRT1}
\end{align}
where $N_\mathrm{on}$ and $R_\mathrm{on}$ have been defined in \eqref{eq:Non_def} and \eqref{eq:Ron_def}, respectively. 
Next, we differentiate  $P(\mathbf{r}(k,L)|\mathbf{m}(k,L),\hat n_r, n_b)$   with respect to $n_b$; and then, similarly, the solution of equation \eqref{eq:diff1} is obtained as 
\begin{equation}
n_b = \hat n_b (\mathbf m(k,L)) = \frac{R_\mathrm{off}(\mathbf{m}(k,L))}{N_\mathrm{off}(\mathbf{m}(k,L))}.
\label{eq:nb_estimate}
\end{equation}
In \eqref{eq:nb_estimate}, $N_\mathrm{off}(\mathbf m(k,L))$ and  $R_\mathrm{off}(\mathbf m(k,L))$ are   functions of $(\mathbf m(k,L))$, and they are defined as 
\begin{align}
N_\mathrm{off}(\mathbf{  m}(k,L))&=\sum_{i=k-L+1}^k (1-m(i)), 
\label{eq:Noff_def}
\end{align}
and
\begin{align}
R_\mathrm{off}(\mathbf{  m}(k,L))&=\sum_{i=k-L+1}^k (1-m(i))r(i).
\label{eq:Roff_def}
\end{align}
Similar to $N_\mathrm{on} $ and  $R_\mathrm{on} $, for simplicity of notation, we also drop the dependence on  $\mathbf m(k,L)$ and use $N_\mathrm{off} $ and  $R_\mathrm{off} $ instead. 
After substituting the solution $n_b = \hat n_b(\mathbf{m}(k,L))$ back into $P (\mathbf{r}(k,L)|\mathbf{m}(k,L),\hat n_r, n_b)$ and eliminating  irrelevant terms, we obtain the decision metric  
\begin{align}
\label{eq:metric_glrt_seq_temp}
\lambda'_2(\mathbf{m}(k,L)) = 
\left(\frac{R_\mathrm{off}}{N_\mathrm{off}}\right)^{R_\mathrm{off}}  
\left(\frac{R_\mathrm{on}}{N_\mathrm{on}}\right)^{R_\mathrm{on}}. 
\end{align}
However, in simulation, we observed that the value of some parts in the metric \eqref{eq:metric_glrt_seq_temp} would become too large and cause a memory overflow problem on the computer (larger than $10^{500}$). 
Hence, by taking $\ln ()$ of the right side of \eqref{eq:metric_glrt_seq_temp}, we obtain our GLRT sequence receiver's decision metric 
\begin{align}
&\lambda_2(\mathbf{m}(k,L))   = \ln (\lambda'_2(\mathbf{m}(k,L))) \nonumber \\
 =& 
 {R_\mathrm{off}}  \ln
\left(\frac{R_\mathrm{off}}{N_\mathrm{off}}\right)  + 
{R_\mathrm{on}} \ln 
\left(\frac{R_\mathrm{on}}{N_\mathrm{on}}\right).
\label{eq:metric_glrt_seq} 
\end{align}
Similar to \eqref{eq:receiver_mlsd} and \eqref{eq:receiver_gmlsd_seq}, the decision of this GLRT sequence receiver is made by performing 
\begin{align}
\mathbf{\hat{m}}(k,L) = \arg \max_{\mathbf{m}(k,L)} \lambda_2(\mathbf{m}(k,L)).
\label{eq:receiver_glrt_seq}
\end{align}
 It should be noted that when the observation window size is very small and the background radiation is very weak, i.e., the $n_b$ value is very small, $R_\mathrm{off}$ is very likely to be zero. 
 When $R_\mathrm{off}=0$, for both receiver hardware and simulation software, there might be an ambiguity in evaluating the term ${R_\mathrm{off}}  \ln
\left(\frac{R_\mathrm{off}}{N_\mathrm{off}}\right)$ of the decision metric \eqref{eq:metric_glrt_seq}. 
Thus, we use the limit of that term when $R_\mathrm{off}\rightarrow0$ to define its actual value at $R_\mathrm{off}=0$, i.e., in implementation, we pre-define
 \begin{align}
 \left. {R_\mathrm{off}}  \ln
\left(\frac{R_\mathrm{off}}{N_\mathrm{off}}\right)  \right|  _{R_\mathrm{off} = 0} 
= \lim_{R_\mathrm{off}\rightarrow 0}    {R_\mathrm{off}}  \ln
\left(\frac{R_\mathrm{off}}{N_\mathrm{off}}\right),
 \end{align}
which can easily be shown to be zero.

Clearly, no integrals are involved in the GLRT sequence receiver decision metric \eqref{eq:metric_glrt_seq}, resulting in low computational complexity. It does not require any knowledge of the fading distribution, and therefore is robust and practical.  
Additionally, since the value of $n_b$ is not required in \eqref{eq:metric_glrt_seq}, compared to the GMLSD receiver proposed in \cite{Uysal2010GMLSD}, our GLRT sequence receiver is more practical for implementation. 
In a later section, we will show it is robust in slowly time-varying environments, regardless of the distributions of $h$ and the value of $n_b$.

\subsection{Implementation}

\begin{figure}
\centering 
\includegraphics{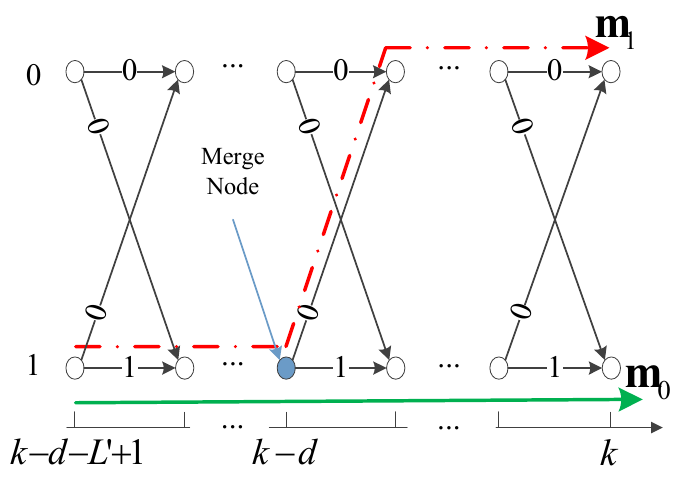}%
\caption{Two survivors in a trellis diagram}
\label{fg:GLRT_trellis}
\end{figure}

 \begin{figure}
\centering 
	\includegraphics{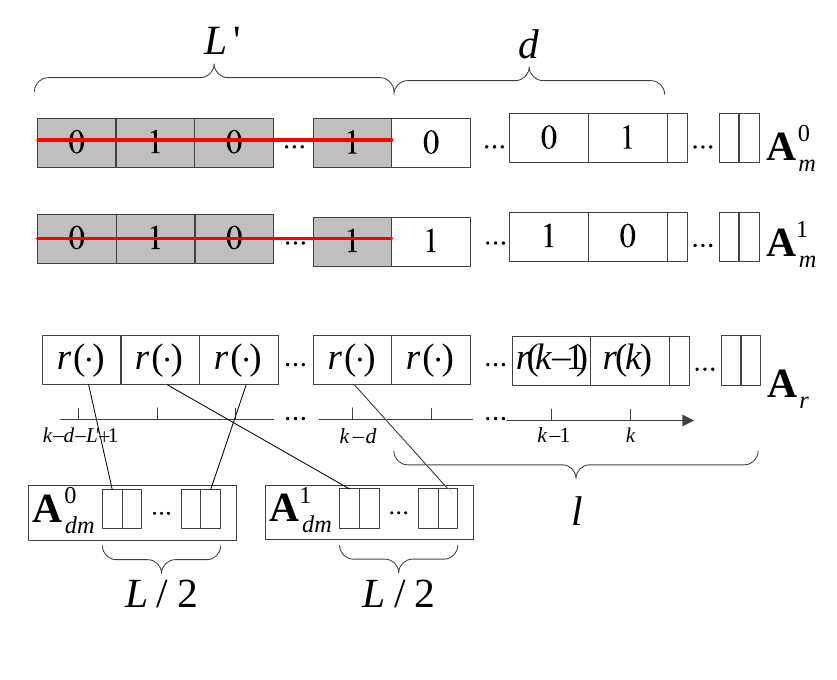}%
	\caption{Memory usage with the Selective-Store Strategy of the GLRT sequence receiver}
	\label{fg:memory_glrt_seq}
\end{figure}

In implemention, we adopt the Viterbi-type trellis-search algorithm and the selective-store strategy similar to what we have introduced in \cite{Song2014GlobeCom, Song2014Arobust} for implementing the decision metric \eqref{eq:metric_glrt_seq} that we use for selection of survivors. 
A trellis diagram is shown in Fig. \ref{fg:GLRT_trellis}, where there are two nodes at each time $k$ and each node is labelled corresponding to data symbol ``0'' and ``1''.
For each node, there are two paths entering it and the Viterbi-type trellis-search algorithm keeps the one with a higher metric value and discards the other. 
Thus, at each time $k$, only two paths exist as survivors and the tail before the merge point of the two survivors give the firm output decisions.

The selective-store fundamental is the same as that in \cite{Song2014Arobust} but the specific strategy is modified here.
We name the part before the merge point the \emph{detected} part, and the part after, the \emph{ongoing} part. 
Also, we name signals that are detected to carry data symbol ``1'' the 1-detected signals; and similarly, the 0-detected signal is defined as the signal that is detected to carry data symbol ``0''. 
Clearly, $R_\mathrm{on}$ can be obtained as $R_\mathrm{on} = R_\mathrm{on-detected}+R_\mathrm{on-ongoing}$, and similarly for  $N_\mathrm{on}$, $R_\mathrm{off}$ and $N_\mathrm{off}$.  
In this selective-store strategy, we keep the values of $N_\mathrm{on-detected}$ and $N_\mathrm{off-detected}$ the same, i.e., $N_\mathrm{on-detected} = N_\mathrm{off-detected} = L/2$. 
As shown in Fig. \ref{fg:memory_glrt_seq}, there are two memory arrays $\mathbf{A}^0_{dr}$ and $\mathbf{A}^1_{dr}$ in the receiver. 
We selectively store the most recent $L/2$ 0-detected signals in $\mathbf{A}^0_{dr}$, and the most recent $L/2$ 1-detected signals in $\mathbf{A}^1_{dr}$.
In operation, when a new 1-detected (0-detected) signal is detected, the system drops the oldest 1-detected (0-detected) signal stored in $\mathbf{A}^1_{dr}$ ($\mathbf{A}^0_{dr}$) and puts the new signal in. 
In this way, the values of $R_\mathrm{on-detected}$ and $R_\mathrm{off-detected}$ can be calculated by recursively subtracting the oldest and adding the newest. 
For the on-going part, we use arrays $\mathbf{A}^0_{om}$ and $\mathbf{A}^1_{om}$ to store the two survivors and array $\mathbf{A}^0_{or}$ to store the undetected signals.
The values of $ R_\mathrm{on-detected}$, $ R_\mathrm{ffn-detected}$, $N_\mathrm{on-detected}$ and $N_\mathrm{off-detected}$ are calculated based on the values stored in arrays $\mathbf{A}^0_{om}$,   $\mathbf{A}^1_{om}$ and $\mathbf{A}^0_{or}$. 
In Fig. \ref{fg:memory_glrt_seq}, $d$ and $l$ are the lengths of the sequence ongoing part and the memory array for storing the ongoing part. 
Apparently, $d$ is a random variable.
In simulations, we observe that the average value of $d$ is smaller than 3 and to ensure $d<l$, we set $l=30$ in implementation. 
We can see that the metric evaluation complexity with our selective-store strategy is still very low and independent of $L$.

After adopting the selective-store strategy, the receiver is probably not using a subsequence with consecutive signals to form the decision metric \eqref{eq:metric_glrt_seq} because it is with a high probability that there are unequal numbers of 0-detected signals and 1-detected signals in a subsequence with length $L$.
We use $L'$ to denote the length of the effective subsequence, in which $\min \{\# \text{ of zeros}, \# \text{ of ones}\}=L/2$.
Thus, the whole observation window length $L_w$ is   $L_w = L' + d$.
In order to perform robust data detection, we should ensure $L_w \ll L_c$.
Apparently, $L_w$ is of the same order of magnitude of $L$ and since $L \ll L_c$, we have $L_w \ll L_c$.

\section{The DFB  Receivers}

\subsection{The GLRT DFB Receiver}

In this subsection, we propose a DFB symbol-by-symbol receiver to further reduce the implementation complexity. 
We use $\hat{m}(k)$ to denote the decision result at time $k$, and $\mathbf{\hat m}(k-1;L) = [ \hat{m}(k-L), ... , \hat{m}(k-1)]$ to denote the decision result vector at time $k-1$ with length $L$.


As symbol-by-symbol detection is performed here, at time $k$, all the detection results before $k$ should be available at the receiver side. 
Thus, when detecting the $k$th symbol, we can consider two hypothesis sequences, which are $[\mathbf{\hat m}(k-1,L),1]$ and $[\mathbf{\hat m}(k-1,L),0]$. 
By comparing the two corresponding decision metrics, and discarding the one with smaller value, we can make the decision.
Formally, the detection at time $k$ can be made by performing 
\begin{align}  
\lambda_2([\mathbf{\hat m}(k-1,L),1)
 \mathop{\gtrless} \limits^{\hat m(k) = 1}_{\hat m(k) = 0}     
\lambda_2([\mathbf{\hat m}(k-1,L),0]).
\label{eq:glrt_dfb1}
\end{align}
As this is a symbol-by-symbol detection, trellis-search is not required here; while the selective-store strategy is adopted.
We store the $L$ most recent 1-detected data symbols as well as the $L$ most recent 0-detected data symbols.
In this way, $N_\mathrm{off}$ and $N_\mathrm{on}$ both equal $L$, constantly.
We simplify \eqref{eq:glrt_dfb1} and then obtain
\begin{align}  
\Psi (R_\mathrm{on},R_\mathrm{off},N_\mathrm{on},N_\mathrm{off},r(k)])
 \mathop{\gtrless} \limits^{\hat m(k) = 1}_{\hat m(k) = 0}     0,
\label{eq:receiver_glrt_dfb}
\end{align}
where 
\begin{align}
&\Psi (R_\mathrm{on},R_\mathrm{off},N_\mathrm{on},N_\mathrm{off},r(k)) \nonumber \\
=  & r(k) \ln\left(\frac{N_\mathrm{off}+1}{R_\mathrm{off}+r(k)}  
 \frac{R_\mathrm{on}+r(k)}{N_\mathrm{on}+1}\right)  \nonumber \\
&\quad -   {R_\mathrm{on}} \ln\left(\frac{N_\mathrm{on}+1}{R_\mathrm{on}+r(k)}  \frac{R_\mathrm{on}}{N_\mathrm{on}} \right) \nonumber \\
&\quad \quad - {R_\mathrm{off}}  \ln\left(\frac{R_\mathrm{off}+r(k)} {N_\mathrm{off}+1}\frac{N_\mathrm{off}}{R_\mathrm{off}} \right).  \nonumber \\
\label{eq:metric_glrt_dfb}
\end{align}
Similar to the case of \eqref{eq:metric_glrt_seq}, when $R_\mathrm{off}=0$, we have a problem in evaluating the value of ${R_\mathrm{off}}  \ln  \left(\frac{R_\mathrm{off}+r(k)} {N_\mathrm{off}+1}\frac{N_\mathrm{off}}{R_\mathrm{off}} \right)$. Thus, we define 
 \begin{align}
& {R_\mathrm{off}}  \ln \left.\left(\frac{R_\mathrm{off}+r(k)} {N_\mathrm{off}+1}\frac{N_\mathrm{off}}{R_\mathrm{off}} \right) \right|  _{R_\mathrm{off} = 0} 
\nonumber \\ 
 =& \lim_{R_\mathrm{off}\rightarrow 0} {R_\mathrm{off}}  \ln\left(\frac{R_\mathrm{off}+r(k)} {N_\mathrm{off}+1}\frac{N_\mathrm{off}}{R_\mathrm{off}} \right) = 0.
\end{align}  
 
As this DFB receiver \eqref{eq:receiver_glrt_dfb} is obtained by simplifying \eqref{eq:receiver_glrt_seq}, we call receiver \eqref{eq:receiver_glrt_dfb} the GLRT DFB receiver.
It is obvious that the implementation complexity of receiver \eqref{eq:receiver_glrt_dfb} is lower than that of receiver \eqref{eq:receiver_glrt_seq}, because each time only one metric is required to be evaluated and no searching is performed.   
In later sections, we will compare the performance of the two receivers.
In the same way,  based on \eqref{eq:receiver_gmlsd_seq}, we develop a DFB receiver, which can be seen as a special case of  \eqref{eq:receiver_glrt_dfb}.

In Appendix A of this paper, we analytically show that as  $L$ increases, the BEP of the GLRT DFB receiver can approach the Genie Bound.
Since the GLRT DFB receiver is a simplified version of the GLRT sequence receiver, apparently, so does the GLRT sequence receiver.  

\subsection{A Special Case - The GMLSD DFB receiver}
Similar to \eqref{eq:glrt_dfb1}, detection can also be made by performing 
\begin{align}  
\lambda_1([\mathbf{\hat m}(k-1,L),1])
 \mathop{\gtrless} \limits^{\hat m(k) = 1}_{\hat m(k) = 0}     
\lambda_1([\mathbf{\hat m}(k-1,L),0]). 
\label{eq:gmlsd_dfb1}
\end{align}
We store the $L$ most recent 1-detected data symbols.
After simplifying \eqref{eq:gmlsd_dfb1}, we obtain
\begin{align}  
\Psi_0 (R_\mathrm{on},R_\mathrm{off},r(k))
 \mathop{\gtrless} \limits^{\hat m(k) = 1}_{\hat m(k) = 0}     0,
\label{eq:receiver_gmlsd_dfb}
\end{align}
where 
\begin{align}
&\Psi_0 (R_\mathrm{on},R_\mathrm{off},r(k)) \nonumber \\
=& (R_\mathrm{on}+r(k)) \ln \left(\frac{R_\mathrm{on}+r(k)}{(N_\mathrm{on}+1) n_b} \right) \nonumber   \\ 
&   - R_\mathrm{on} \ln \left(\frac{R_\mathrm{on}}{N_\mathrm{on}n_b} \right)    -r(k)+n_b   .
\end{align}
 
Similarly, this receiver is called the GMLSD DFB receiver and it is with lower implementation complexity compared to \eqref{eq:receiver_gmlsd_seq}.  

\section{Simulations and Results}

As mentioned in section II, in this paper we only use the Gamma-Gamma distribution to model $h_a$. 
For the weak turbulence channel, parameters are set as  $\alpha = 17.13$, $\beta = 16.04$ and the corresponding $\text{SI}=0.1244$; and for the strong turbulence channel, $\alpha=2.23$, $\beta=1.54$, and $\text{SI}=1.3890$. 
We choose parameters $A_0=0.0198$ and $\gamma^2=2.8071$ for \eqref{eq: pdf_pointing} for the pointing errors.
Similar to \cite{Schober2008PCTWC}, without loss of generality, the pdf of $h$ has been normalized such that the mean channel gain is unity, i.e. $\mathbb{E}[h]=1$. 
In all the figures, each curve with a legend  ``num.'' is  obtained by numerical integration; and the one with a legend ``sim.'' is obtained by simulation.

For different SI values (0.1244 and 1.3890) and different $n_b$ values (from 5 to 100), the simulation results obtained by implementing the PCSI receiver \eqref{eq:receiver_ideal} are in complete accord with the corresponding results by numerically integrating the BEP formula \eqref{eq:bep_ideal}.
Hence, in this paper, when referring to the performance of the PCSI receiver, we only give the numerical integration results.

\subsection{Sequence Receivers}

\begin{figure} 
\subfigure[$n_b=39$.]{  
\includegraphics[scale=0.96]{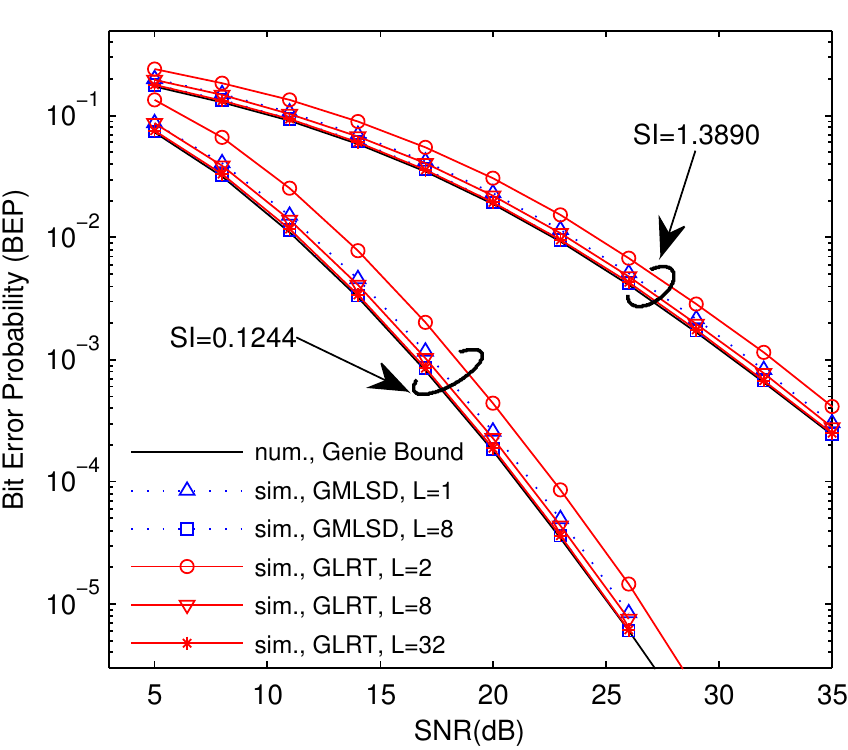}
\label{fg:seq_snr_39}
}
\subfigure[$n_b=20$.]{
\includegraphics[scale=0.96]{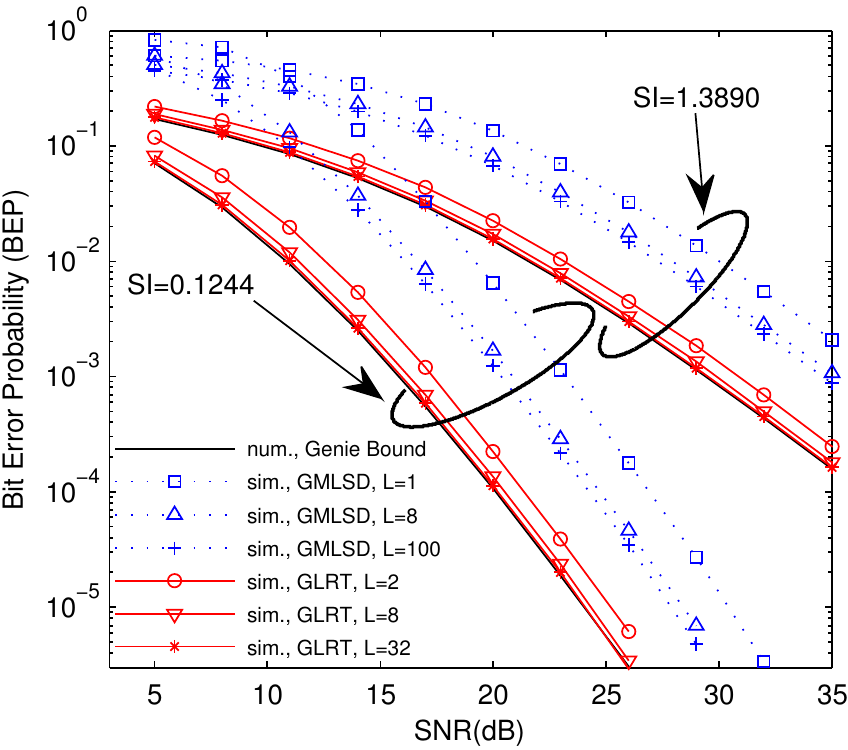}
\label{fg:seq_snr_20}
} 
\subfigure[$n_b=60$.]{
\includegraphics[scale=0.96]{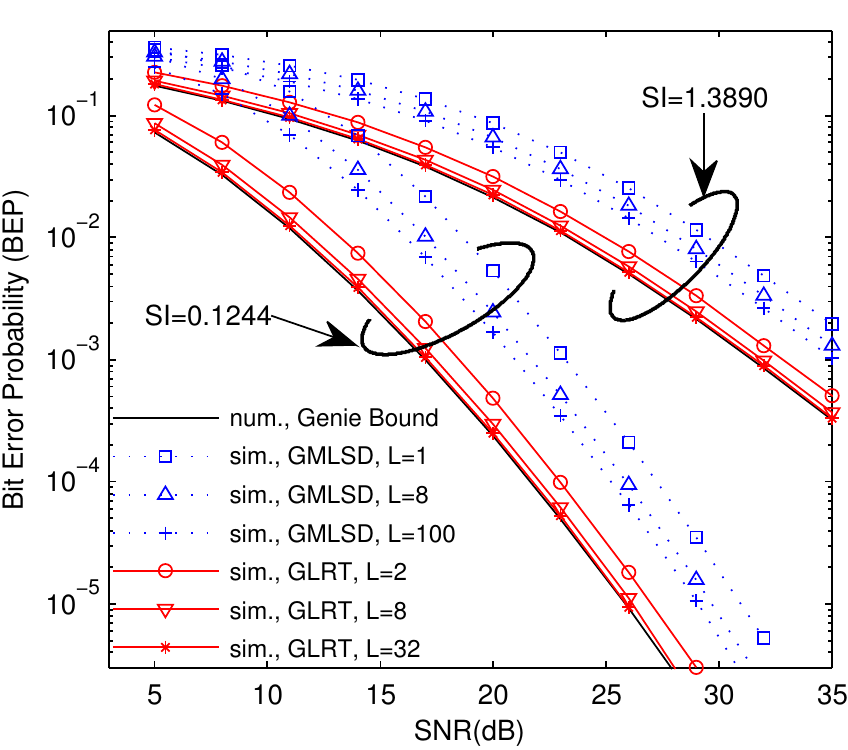}
\label{fg:seq_snr_60}
} 
\caption{Performance of sequence receivers.}
\label{fg:seq_snr}
\end{figure}

In Fig. \ref{fg:seq_snr}, we plot the performance of the GMLSD sequence and the GLRT sequence receiver with different $n_b$ values. 
Both receivers are implemented with the Viterbi-type trellis-search algorithm with the selective-store strategy. 
In Fig. \ref{fg:seq_snr_39} where $n_b=39$, we can see that when $L=1$, the power loss of the GMLSD sequence receiver (with the precise knowledge of $n_b=39$) compared to the PCSI receiver is very small, approximately 0.3 dB; and the power loss of the GLRT sequence receiver when $L=2$ is slightly larger, at approximately 1 dB. 
When the value of $L$ increases to 8, the power loss of the GMLSD sequence receiver cannot be observed; i.e., the performance achieves that of the PCSI receiver. 
For the GLRT sequence receiver, to achieve the Genie Bound, $L$ needs to be no less than 32.   
This is because the GLRT sequence receiver has to estimate two channel parameters ($n_r$ and $n_b$), but the GMLSD receiver only estimates one ($n_r$). 
When the system memory length increases, both sequence receivers can estimate unknown channels almost perfectly and thus achieve the Genie Bound.  
We are interested in how the two sequence receivers will perform if $n_b$ is not perfectly known at the receiver.

In Fig. \ref{fg:seq_snr_20} where $n_b=20$ and Fig. \ref{fg:seq_snr_60} where $n_b=60$, we show the simulation results for both receivers.
Similar to that in Fig. \ref{fg:seq_snr_39}, our GLRT sequence receiver suffers from an approximate 1 dB power loss compared to the Genie Bound when $L=1$, and achieves the Genie Bound when $L = 32$.
However, we can see that the performance of the GMLSD receiver deteriorates whenever $n_b$ increases or decreases.
Even with a very large value of $L$, i.e.,  $L=100$, since it does not know that the $n_b$ value has been changed from 39, its performance cannot converge to the Genie Bound.

Since we adopt the Viterbi-type trellis-search algorithm,  we can simulate  these two sequence receivers efficiently with any arbitrary large value of $L$ ($L \ll L_c$) to achieve the optimum performance. 
The adoption of the selective-store strategy helps us efficiently use the memory and completely avoid the potential error floor.

\begin{figure} 
\includegraphics[scale=0.96]{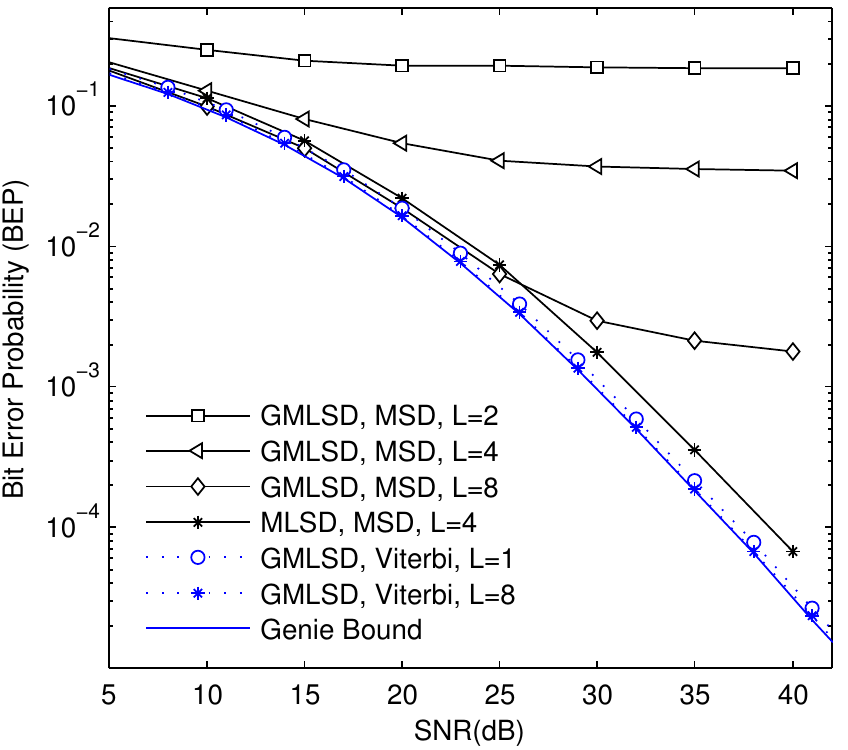}
 \caption{Comparison of the MSD algorithm and the Viterbi-type trellis-search algorithm for the GMLSD sequence receiver; $\alpha=2.23$, $\beta=1.54$, i.e., $\text{SI}=1.3890$, but pointing error is not considered. }
\label{fg:compare}
\end{figure}

When implementing the GMLSD receiver, the MSD algorithm proposed in \cite{Schober2008PCTWC} is adopted by the authors of \cite{Uysal2010GMLSD}, where an error floor is observed with low $L$ values.
We also simulate the GMLSD sequence receiver using our Viteri-type trellis-search algorithm with selective store strategy and plot the results in Fig. \ref{fg:compare}.
We can clearly see that, with our implementation method, the GMLSD sequence receiver performs much better and also completely avoids the potential error floor.   
 
\subsection{DFB Receivers}
 
\begin{figure} 
\subfigure[SI=0.1244.]{  
\includegraphics[scale=0.96]{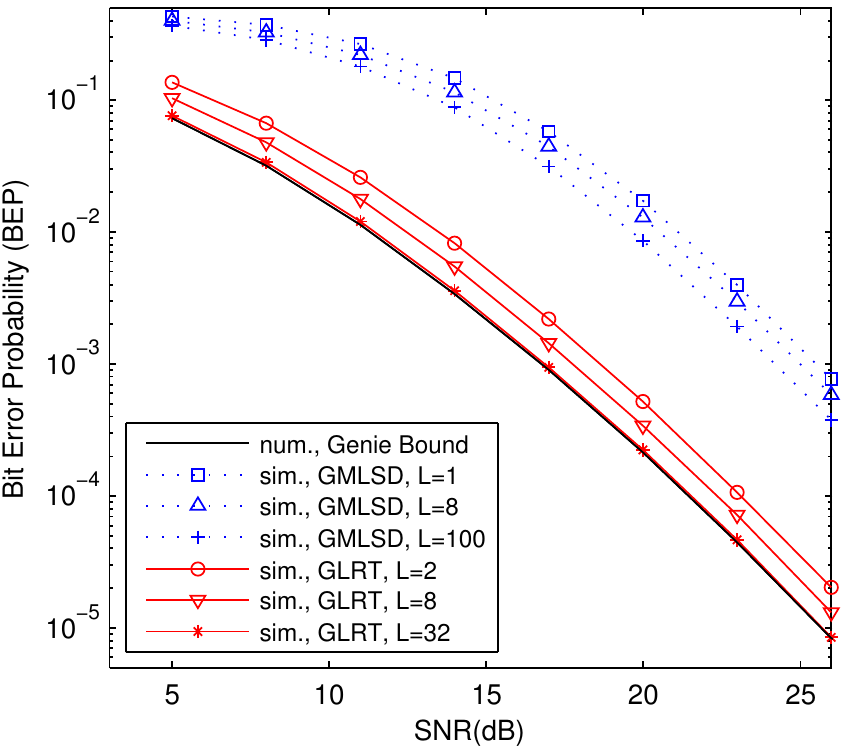}
\label{fg:dfb_snr_w}
}
\subfigure[SI=1.3890.]{
\includegraphics[scale=0.96]{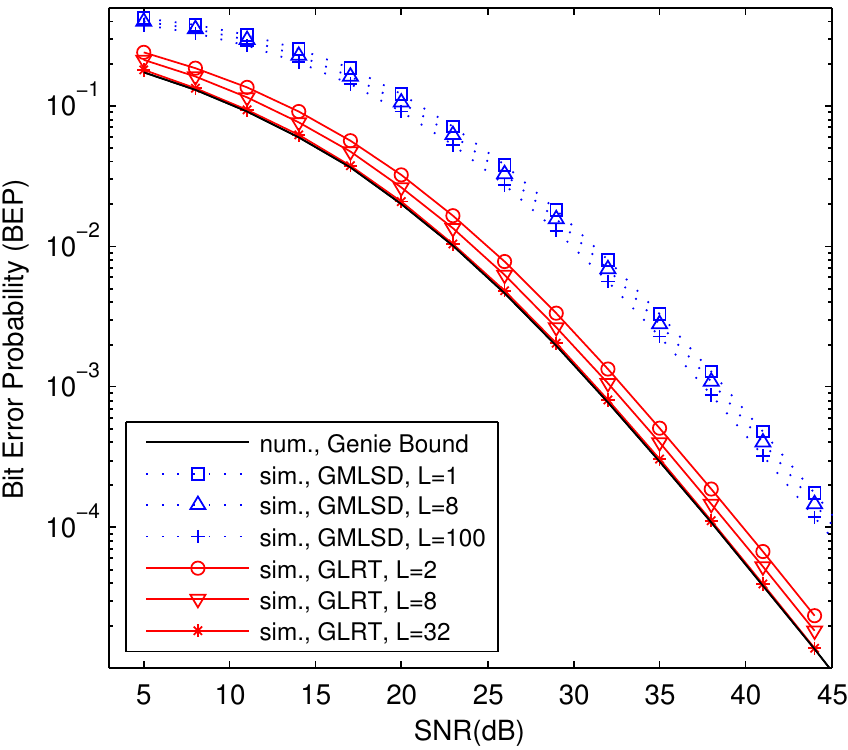}
\label{fg:dfb_snr_s}
}  
\caption{Performance of DFB receivers, where $n_b$ is randomly chosen from 10 to 100.}
\label{fg:dfb_snr_uniform}
\end{figure}

By simulation, using the same channel conditions (turbulence conditions, SNR, $n_b$ value and $L$ value), we find that the DFB receivers have almost the same performance as their corresponding sequence receivers. 
Thus, readers are suggested to refer to Fig. \ref{fg:seq_snr} and we do not replot them.  
Here, we present the DFB receivers' performance where $n_b$ is chosen randomly from 10 to 100 with equal probability. 
Since the background radiation condition (the statistical distribution of the radiation strength) might be different in different situations, in our work here, we just use this randomly chosen $n_b$ to test the robustness of our receivers.  
First, in Fig. \ref{fg:dfb_snr_uniform}, we plot the simulation results of both our GLRT DFB receiver and the GMLSD DFB receiver. 
We can see that our GLRT DFB receiver achieves the Genie Bound with $L=32$, but the GMLSD DFB receiver cannot.

\begin{figure} 
\includegraphics[scale=0.96]{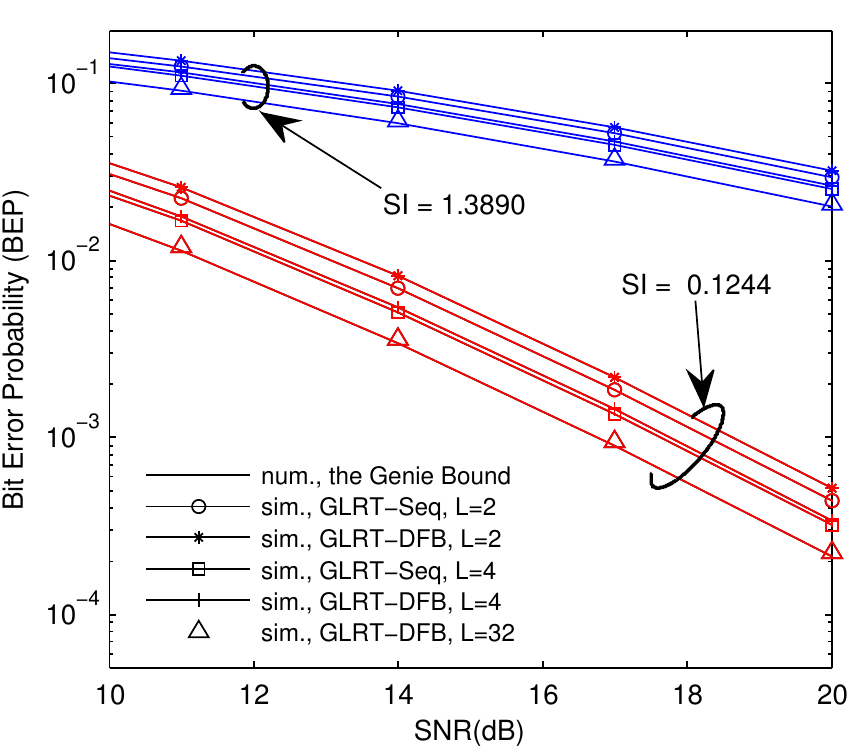}
 \caption{Performance  of both GLRT receivers, where $n_b$ is randomly chosen from 10 to 100.}
\label{fg:zoom_uniform}
\end{figure}

\begin{figure} 
\includegraphics[scale=0.96]{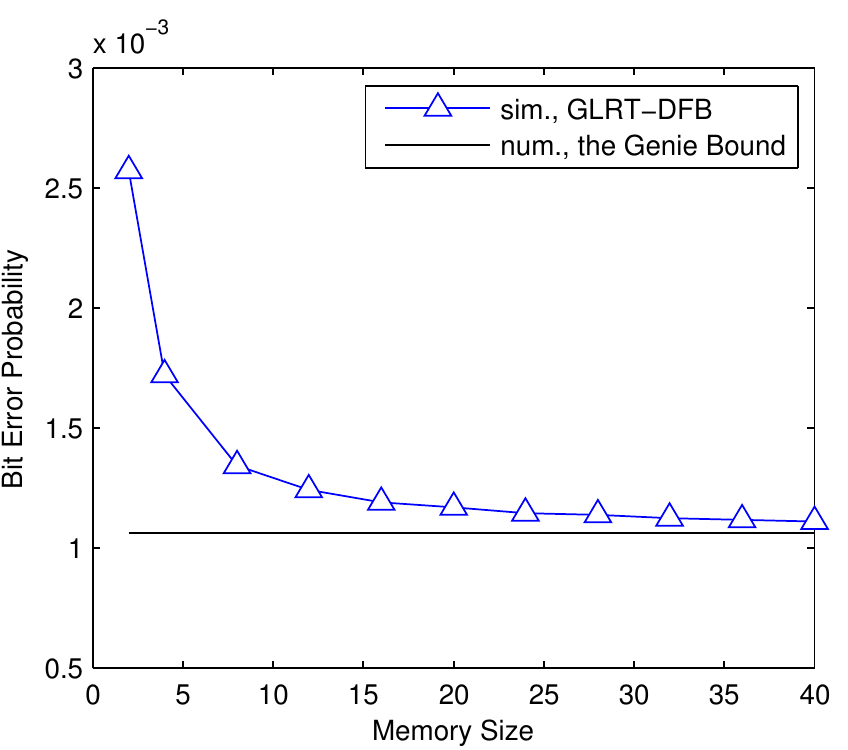}
 \caption{Performance with different memory lengths; SI=0.1244, $n_b=70$, $\mathbf{SNR}=17\text{dB}$}  
\label{fg:dfb_ltest}
\end{figure}

Next, we plot the performance results of the GLRT DFB receiver in Fig. \ref{fg:zoom_uniform} compared to the GLRT sequence receiver. 
We can see that the GLRT sequence receiver performs slightly better than the  GLRT DFB receiver  when $L=2$. 
When $L$ increases, the BEP of the GLRT DFB receiver can also approach Genie Bound. 
 
In Fig. \ref{fg:dfb_ltest}, we plot the GLRT DFB receiver's performance with different values of $L$.  
It can be seen that, as $L$ increases, the performance approaches the Genie Bound asymptotically.

\subsection{Discussion}
Based on the results shown in the previous subsection, we have shown that sequence receivers can be implemented using the Viterbi-type trellis-search algorithm and the selective-store strategy with lower complexity and better performance compared to the MSD algorithm.
The error floor problem, which is observed in \cite{Uysal2010GMLSD} and cannot be mitigated by the MSD algorithm, is now overcome completely.

When $n_b$ is a constant and known at the receiver side and the value of $L$ is small, the GLRT receivers cannot perform as well as the GMLSD receivers, because the GLRT receivers always regard the value of $n_b$ as unknown and have to estimate it with insufficient samples.  
Thus, the GMLSD receivers are suggested only when the receiver memory size is a constraint and the value of $n_b$ is constant and accurately known at the receiver side. 
Otherwise, the GLRT receivers, with more robust performance and require no knowledge of $n_b$, are highly suggested.   

Though the DFB receivers do not perform as well as their corresponding sequence receivers at the same value of $L$, the DFB receivers are recommended because of much lower memory requirements.
We always have to preserve several memory arrays for the sequence receivers to store the ongoing part of survivors.
Besides, the sequence receivers have a potential to introduce a long system delay when the two survivors are very long before they merge. 
On the contrary, since the DFB receivers perform symbol-by-symbol detection, memory arrays to store survivors are not necessary and no system delay exists.
In addition, we observed that by increasing the value of $L$ very slightly, the DFB receiver can perform better than the sequence receiver.

\section{Concluding Remarks}

To mitigate the effects of atmospheric turbulence, pointing errors and background radiation, we have introduced new GLRT receivers (the GLRT sequence receiver and the GLRT DFB receiver) for FSO photon counting systems. 
These GLRT receivers both can perform ML estimation of the unknown channel gain and background radiation implicitly, and detect the data accordingly, while requiring no prior knowledge about the channel and the environment. 
Thus, they are robust and work well in any slowly time-varying environment. 
Using the Viterbi-type trellis-search algorithm as well as the selective-store strategy, the GLRT sequence receiver can be implemented efficiently; while the GLRT DFB receiver is a more efficient option. 
By simulation and theoretical analysis, we have shown that the performance of the GRT receivers approaches the Genie Bound as the observation window length increases.

When the background radiation is constant and perfectly known at the receiver side, the performance of the GMLSD receivers (the GMLSD sequence receiver and the GMLSD DFB receiver) can also achieve the Genie Bound as the observation window length increases. 
By adopting the Viterbi-type trellis-search algorithm and the selective-store strategy, for both GLRT and GMLSD sequence receivers, the error floor problem has been completely avoided. 

Additionally, in Appendix A of this article, we argue that the intuitive detection method will have an error floor problem, and thus is not recommended.

\section*{Appendix A}

From \eqref{eq:receiver_glrt_dfb}, we know that the GLRT DFB receiver decision rule is \begin{multline}
 r(k) \ln\left[\frac{N_\mathrm{off}+1}{R_\mathrm{off}+r(k)}  
 \frac{R_\mathrm{on}+r(k)}{N_\mathrm{on}+1}\right]
\\
 \mathop{\gtrless} \limits^{\hat m(k) = 1}_{\hat m(k) = 0}   \\
  {R_\mathrm{on}} \ln\left[\frac{N_\mathrm{on}+1}{R_\mathrm{on}+r(k)}  \frac{R_\mathrm{on}}{N_\mathrm{on}} \right]
+{R_\mathrm{off}}  \ln\left[\frac{R_\mathrm{off}+r(k)} {N_\mathrm{off}+1}\frac{N_\mathrm{off}}{R_\mathrm{off}} \right].
\label{eq:perf_1}
\end{multline}
The right hand side of \eqref{eq:perf_1} can be expressed as
\begin{align}
&  {R_\mathrm{on}} \ln\left[\frac{N_\mathrm{on}+1}{R_\mathrm{on}+r(k)}  \frac{R_\mathrm{on}}{N_\mathrm{on}} \right]   
 +{R_\mathrm{off}}  \ln\left[\frac{R_\mathrm{off}+r(k)} {N_\mathrm{off}+1}\frac{N_\mathrm{off}}{R_\mathrm{off}} \right]  \nonumber \\
= &R_\mathrm{on}\ln\left[ (1+\frac{1}{N_\text{on}}) (1-\frac{r(k)}{R_\text{on}+r(k)}) \right] \nonumber \\
&\qquad +R_\mathrm{off} \ln\left[ (1+\frac{r(k)}{R_\text{off}})(1-\frac{1}{N_\text{off}+1}) \right]\nonumber  \\
=&  R_\mathrm{on} \ln \left[ 1+ \frac{1}{N_\text{on}} -\frac{r(k)}{R_\text{on}+r(k)} -\frac{1}{N_\text{on}} \frac{r(k)}{R_\text{on}+r(k)}  \right] \nonumber  \\
&  \qquad   + R_\mathrm{off} \ln \left[ 1+ \frac{r(k)}{R_\text{off}} -\frac{1}{N_\text{off}+1} -\frac{r(k)}{R_\text{off}}\frac{1}{N_\text{off}+1}  \right]. 
\end{align}
If  $L$ goes to infinity, we have $R_\mathrm{on}\rightarrow \infty , R_\mathrm{off}\rightarrow \infty , N_\mathrm{on}\rightarrow \infty , \text{ and }N_\mathrm{off} \rightarrow \infty $.
Since $\lim_{x\rightarrow 0}\ln(1+x) = x$, we have that if $L \rightarrow \infty$
\begin{align}
  \frac{1}{N_\text{on}} -\frac{r(k)}{R_\text{on}+r(k)} -\frac{1}{N_\text{on}} \frac{r(k)}{R_\text{on}+r(k)} \rightarrow 0,
\end{align}
and
\begin{align}
 \frac{r(k)}{R_\text{off}} -\frac{1}{N_\text{off}+1} -\frac{r(k)}{R_\text{off}}\frac{1}{N_\text{off}+1}  \rightarrow 0.
\end{align}
Thus, we have 
\begin{align}
& \lim_{L \rightarrow \infty}    R_\mathrm{on} \ln \left[ 1+ \frac{1}{N_\text{on}} -\frac{r(k)}{R_\text{on}+r(k)} -\frac{1}{N_\text{on}} \frac{r(k)}{R_\text{on}+r(k)}  \right] \nonumber  \\
&  \qquad   + R_\mathrm{off} \ln \left[ 1+ \frac{r(k)}{R_\text{off}} -\frac{1}{N_\text{off}+1} -\frac{r(k)}{R_\text{off}}\frac{1}{N_\text{off}+1}  \right] \nonumber \\
=&  \frac{R_\mathrm{on}}{N_\text{on}}  - \frac{R_\mathrm{off} }{N_\text{off}+1} \nonumber  \\
&+r(k) \left[ 1 - \frac{R_\text{on}  }{R_\text{on}+r(k)}(1+ \frac{1}{N_\text{on}})  - \frac{1}{N_\text{off}+1}  \right]. \nonumber  \\
\end{align}
Furthermore, we have
\begin{align}
&\lim_{L \rightarrow \infty} \bigg( \frac{R_\mathrm{on}}{N_\text{on}}  - \frac{R_\mathrm{off} }{N_\text{off}+1} \nonumber  \\
&+r(k) \left[ 1 - \frac{R_\text{on}  }{R_\text{on}+r(k)}(1+ \frac{1}{N_\text{on}})  - \frac{1}{N_\text{off}+1}  \right] \bigg) \nonumber \\
= & \frac{R_\mathrm{on}}{N_\text{on}}  - \frac{R_\mathrm{off} }{N_\text{off}},
\end{align}
and
\begin{align}
\lim_{L \rightarrow \infty} \frac{N_\mathrm{off}+1}{R_\mathrm{off}+r(k)}  
 \frac{R_\mathrm{on}+r(k)}{N_\mathrm{on}+1}  = 
 \frac{N_\mathrm{off}}{R_\mathrm{off}}  
 \frac{R_\mathrm{on}}{N_\mathrm{on}}. 
\end{align}
Finally, we can see that \eqref{eq:perf_1}   can be reduced to  
\begin{align}
  r(k) \ln\left[  \frac{N_\mathrm{off}}{R_\mathrm{off}}  
 \frac{R_\mathrm{on}}{N_\mathrm{on}} \right]
 \mathop{\gtrless} \limits^{\hat m(k) = 1}_{\hat m(k) = 0}    
\frac{R_\mathrm{on}}{N_\text{on}}  - \frac{R_\mathrm{off} }{N_\text{off}}
\label{eq:perf3}
\end{align}
If we substitute \eqref{eq:nb_estimate} into \eqref{eq:nr_estimate}, we have the ML estimate of $n_r$ as 
\begin{align}
\hat{n}_r = \frac{R_\mathrm{on}}{N_\text{on}}  - \frac{R_\mathrm{off} }{N_\text{off}}.
\label{eq:n_r_ml}
\end{align}
Substituting \eqref{eq:nb_estimate} and \eqref{eq:n_r_ml} into \eqref{eq:perf3}, we have 
\begin{align}
  r(k)  \mathop{\gtrless} \limits^{\hat m(k) = 1}_{\hat m(k) = 0}  
 \hat{n}_r / \ln \left(1+ \frac{\hat{n}_r }{\hat{n}_b} \right). 
 \label{eq:perf5}
 \end{align}
Comparing \eqref{eq:perf5} with \eqref{eq:receiver_ideal}, we can see that they have the same structure and the only difference is that \eqref{eq:perf5} uses estimated parameter values but \eqref{eq:receiver_ideal} uses ideally accurate ones. 

After adopting the selective-store strategy, we have
\begin{align}
\hat n_b = \frac{2\sum_{i=1}^{L/2} r_{i,k}^0}{L},
\end{align}
where $r_{i,k}^0$ is defined as the $i$th most recent received signal at time $k$ that is detected to carry data symbol 0. 
If we ignore the impact of feedback errors and consider all $r_{i,k}$'s as the received signal corresponding to the data symbol 0, $r_{i,k}^0$ is a random variable with mean $n_b$ and variance $n_b$.
Therefore, $\hat{n}_b$ is a random variable with mean $n_b$ and variance $2n_b/L$. 
Clearly, if $L$ goes to infinity, $2n_b/L$ goes to zero and $\hat{n}_b$ approaches the true value of $n_b$.
Similarly, it can be shown that as  $L$ increases, $\hat{n}_r$  approaches its true value ${n_r}$.
Thus, we can conclude that as the observation window length increases, the BEP of the GLRT DFB receiver can approach the Genie Bound, and apparently, so does the GLRT sequence receiver.

%
%
%



\begin{IEEEbiographynophoto}{Tianyu Song}
 [S'13] was born in Wuchang, Heilongjiang Province, China in 1989. He received his Bachelor of Engineering from Honors School of Harbin Institute of Technology (HIT) in 2011, Harbin, China. From Sep 2009 to Jun 2010, Tianyu was an exchange student in Department of Electrical Engineering of Korea Advanced Institute of Science and Technology (KAIST), Daejon, South Korea. Now he is with Natioanal University of Singapore and working toward the Ph.D. degree supervised by Professor Pooi-Yuen Kam. Tianyu's research interests include Free Space Optical communication, optimal receiver design, stochastic process and algorithm.
\end{IEEEbiographynophoto}
\begin{IEEEbiographynophoto}{Pooi-Yuen KAM}
 [F'10] was born in Ipoh, Malaysia, and educated at the Massachusetts Institute of Technology, Cambridge, Mass., USA where he obtained the S. B., S. M., and Ph. D. degrees in electrical engineering in 1972, 1973, and 1976, respectively.\par
From 1976 to 1978, he was a member of the technical staff at the Bell Telephone Laboratories, Holmdel, N. J., U. S. A., where he was engaged in packet network studies.  Since 1978, he has been with the Department of Electrical and Computer Engineering, National University of Singapore, where he is now a professor.  He served as the Deputy Dean of Engineering and the Vice Dean for Academic Affairs, Faculty of Engineering of the National University of Singapore, from 2000 to 2003. His research interests are in the communication sciences and information theory, and their applications to wireless and optical communications.  He spent the sabbatical year 1987 to 1988 at the Tokyo Institute of Technology, Tokyo, Japan, under the sponsorship of the Hitachi Scholarship Foundation. In year 2006, he was invited to the School of Engineering Science, Simon Fraser University, Burnaby, B.C., Canada, as the David Bested Fellow.\par
Dr. Kam is a member of Eta Kappa Nu, Tau Beta Pi, and Sigma Xi.  Since September 2011, he is a senior editor of the IEEE Wireless Communications Letters. From 1996 to 2011, he served as the Editor for Modulation and Detection for Wireless Systems of the IEEE Transactions on Communications. He currently also serves on the editorial board of PHYCOM, the Journal of Physical Communications of Elsevier. He was elected a Fellow of the IEEE for his contributions to receiver design and performance analysis for wireless communications. He received the Best Paper Award at the IEEE VTC2004-Fall, at the IEEE VTC2011-Spring, and at the IEEE ICC2011.
\end{IEEEbiographynophoto}

\end{document}